\documentstyle[a4wide,12pt,epsf,psfig]{article}
\begin{document}
\begin{flushright}
\baselineskip=12pt
{SUSX-TH-97/12}\\
{hep-th/9707105}\\
August 1997\\
\end{flushright}

\begin{center}
{\LARGE \bf 
THE SUPERSYMMETRIC CP PROBLEM  IN
ORBIFOLD COMPACTIFICATIONS \\}
\vglue 0.45cm
{D.BAILIN$^{\Diamond}$ \footnote
{D.Bailin@sussex.ac.uk},G. V. KRANIOTIS$^{\heartsuit}$ \footnote
 {G.Kraniotis@rhbnc.ac.uk} and A. LOVE$^{\heartsuit}$ \\}
	{$\Diamond$ \it  Centre for Theoretical Physics, \\}
{\it University of Sussex,\\}
{\it Brighton BN1 9QJ, U.K. \\}
\vglue 0.25cm
{$\heartsuit$ \it  Department of Physics, \\}
{\it Royal Holloway and Bedford New College, \\}
{\it  University of London,Egham, \\}
{\it Surrey TW20-0EX, U.K. \\}
\baselineskip=12pt

\vglue 0.35cm
ABSTRACT
\end{center}

{\rightskip=3pc
\leftskip=3pc
\noindent
\baselineskip=20pt
The possibility of spontaneous breaking of $CP$ symmetry by the
expectation values of orbifold moduli is investigated  with particular
reference to $CP$ violating phases in soft supersymmetry breaking
terms. The effect of different mechanisms for stabilizing the
dilaton and the form of the non-perturbative superpotential on the
existence and size of these phases is studied. 
Models with modular symmetries which are subgroups of
$PSL(2,Z)$, as well as the single overall
modulus $T$ case with the full  $PSL(2,Z)$ modular symmetry, are
discussed.
Non-perturbative 
superpotentials involving the absolute modular invariant $j(T)$,
such as may arise from $F$-theory compactifications, are
considered.
 }

\vfill\eject
\setcounter{page}{1}
\pagestyle{plain}
\baselineskip=14pt
	
\section{Introduction}

The concept of {\it symmetry} has been the most useful guiding 
principle in our search for discovering the fundamental laws of
Nature.
However, the understanding of 
{\it symmetry breaking} stands on an equal footing since
a lot of complex phenomena in our cosmos depend on it.
The observed baryon asymmetry of the Universe, if dynamically generated,
requires that $CP$ is violated.
The observed tiny breaking of $CP$ \cite{nekaon}
symmetry is one of the most intriguing problems in particle physics
and it has now been with us for more than 30 years, 
awaiting an
explanation. It therefore constitutes one  of the most promising
directions in the search for new physics beyond the Standard Model.

String theory may provide a new perspective on the longstanding
question of the origin of $CP$ violation. It has been argued
\cite{dine} that there is no explicit $CP$ symmetry breaking in
string theory whether perturbative or non-perturbative.
However, $CP$ violation might arise from complex expectation values
of moduli or other scalars \cite{dine}-\cite
{steve}. In all supergravity
theories, including those derived from string theory, there is the
possibility of $CP$ violating phases in the soft supersymmetry
breaking $A$ and $B$ terms and gaugino masses which are in addition
to a possible phase in the Kobayashi-Maskawa matrix and the 
$\theta$ parameter of QCD; (see for example, \cite{HALL} and references
therein.) In compactifications of string theory,
soft supersymmetry breaking terms can be functions of moduli
such as those associated with the radius and angles characterizing the
underlying torus of the orbifold compactification. Then, if
these moduli develop complex vacuum expectation values this can
feed through to the low energy supergravity as $CP$ 
violating phases.

Indeed, any nonzero value for $d_n$,
the neutron electric dipole
moment, is an indication of $CP$ violation.
In principle, complex soft superymmetry breaking terms in supersymmetric
(SUSY) theories
and their resulting phases
can lead to large contributions to $d_n$
\cite{josef}. These  phases are constrained
by experiment to be $\leq O(10^{-3})$. It is therefore a serious
challenge for SUSY theories to explain why these phases are so small,
i.e why soft susy breaking terms preserve $CP$ to such a high degree
\cite{nir}.
As we will show below, string supersymmetric theories relate the
required smallness of $CP$ phases to properties of modular functions.

To estimate the size of such $CP$ violating phases for orbifold
compactifications, it is first necessary to minimize the effective
potential to determine the expectation values of the moduli 
fields $T_i$.
Such calculations may be sensitive to the solution proposed
to the problem of stabilizing the dilaton expectation value.
Two mechanisms for stabilizing the dilaton have been proposed. The first
one assumes that more than one gaugino condensate 
\cite{DIXON} is present in the
superpotential $W_{np}$. This class of models 
has been termed as multiple gaugino 
condensate or racetrack models.
The second proposal is more {\it stringy} in nature
\cite{BANK,casa}. 
There are good reasons to believe that 
stringy non-perturbative corrections
to the  {non-holomorphic} K$\rm{\ddot{a}}$hler potential are
sizeable and can stabilize the dilaton, thereby solving the runaway problem.

Modular invariance of the
effective Lagrangian strongly restricts the form of the non-perturbative
superpotential $W_{np}$ and connects it to the theory of modular 
forms \cite{font}. Recently, it has been demonstrated
that superpotentials with modular properties may also arise in 
$F$-theory compactifications
\cite{witten}. 
As has been shown in \cite{
bagger}, the superpotential $W$ is a section of a holomorphic line bundle. 
Among other ways, theta 
functions can be
 viewed as sections of line bundles on abelian varieties and/or the 
moduli space of abelian varieties \cite{mumford}. In $F$-theory constructions
\cite{witten} non-trivial non-perturbative superpotentials involve 
theta functions, in fact an $E_8$ theta function
in the example of \cite{witten}, and as such they 
exhibit definite modular properties. 
These recent developements enhance the expectations that the superpotential
is indeed a {\it modular form} under duality transformations.
In this work we investigate 
the effect that modular invariance and dilaton-stabilization
by the 
mechanisms described above
have on the $CP$ properties of the resulting soft supersymmetry-breaking 
terms.  
 
The material of this paper is organized as follows.
In section 2 we describe the form of the effective potential
that encompasses both possibilities for the stabilization of the
dilaton, and present the soft supersymmetry-breaking terms emerging
from such a potential.
In section 3 we study the case of the $Z6-IIb$ orbifold with 
modular symmetries for some of its moduli 
which are subgroups of $PSL(2,Z)$, and assuming
that the dilaton is stabilised by a multiple gaugino condensate.
In section 4 we study the case of the $Z6-IIb$ orbifold where
non-perturbative stringy corrections to the 
K$\rm{\ddot{a}}$hler potential  are responsible for the stabilization
of the dilaton and present our results. 
In section 5 the case of the overall modulus $T$ 
with a full $PSL(2,Z)$ modular symmetry is presented.
Our conclusions are presented in section 6.

\section{Effective potential and soft supersymmetry breaking terms}

To estimate the size of $CP$ violating phases in the soft supersymmetry-breaking 
terms, we need the form of the effective potential $V_{eff}$.
We can then minimize to find the expectation values of the $T$ moduli,
which are in general complex. The outcome may be sensitive to the
solution proposed to the problem of stabilizing the dilaton expectation
value at a realistic value of $\rm{Re} S$. The possibilities we shall
consider are that this stabilization is due to a multiple gaugino
condensate in the hidden sector \cite{DIXON} (including hidden sector
matter), or due to stringy non-perturbative corrections to the 
K$\rm{\ddot{a}}$hler potential \cite{BANK,casa}.
For convenience, we write down expressions for the effective potential
and for the soft supersymmetry breaking terms that are general 
enough to encompass both possibilities.

The general form of the multiple gaugino condensate non-perturbative 
superpotential $W_{np}$ derived from orbifold compactifications is
\cite{font,dbalws}
\begin{eqnarray}
W_{np}&=& \sum_{_a} h_{a}\; e^{\frac{24 \pi^2 S}{b_a}} \prod_{i,m} \Bigl(
\eta\Bigl(\frac{T_i}{l_{im}}\Bigr)\Bigr)^{-c_{im}(1-\frac{3 \delta_{GS}^i}{b_a})} 
\nonumber \\
& \times & \prod_{j} H_{j}(T_j)
\label{superp}
\end{eqnarray}
for some coefficients $h_a$, where $\delta_{GS}^{i}$ are Green-Schwarz
parameters and $b_a$ are renormalization group coefficients for
the various factors of the hidden sector gauge group. The coefficients
$c_{im}$ and $l_{im}$, which may be found tabulated elsewhere
\cite{dbalws}, characterize the subgroups of the modular symmetry
group $PSL(2,Z)$ that arise for non $T_2+T_4$ orbifolds
\cite{mayr} i.e. orbifolds for which there are some twisted sectors
with fixed planes for which the six-torus $T_6$ can not be
decomposed into a direct sum $T_2+T_4$ with the fixed plane lying
in $T_2$. In general,
\begin{equation}
\sum_{m} c_{im}=2
\label{dyo}
\end{equation}
In (\ref{superp}) the moduli $T_i$ are to be understood to include
the $U$ moduli associated with $Z_2$ planes.
Factors $H_{i}(T_i)$ depending on the absolute modular invariant
$j(T_i)$ have been introduced. Non-perturbative superpotentials
involving $j(T_i)$ may, in principle, arise from orbifold 
theories containing gauge non-singlet states which become 
massless at some special values of the moduli \cite{miriam},
 though examples are lacking. They may also arise from
$F$-theory compactifications \cite{witten}. When the
 modular symmetry associated with $T_i$ is the full 
$PSL(2,Z)$, the most
general form of $H_i(T_i)$ to avoid singularities in the
fundamental domain \cite{miriam} is
\begin{equation}
H_i(T_i)=(j(T_i)-1728)^{m_i/2}\;j(T_i)^{n_i/3} P_i(j(T_i))
\label{Joseph}
\end{equation}
where $m_i$ and $n_i$ are integers and $P_i$ is a polynomial
in $j$. For generic orbifold compactifications the $H_i(T_i)$
are all 1.
Since we do not wish to commit ourselves to any particular choice
of gaugino condensates nor of hidden sector matter, we find it 
convenient to rewrite (\ref{superp}) in the form
\begin{equation}
W_{np}=\Omega(\Sigma) F(T_i)
\label{sigma}
\end{equation}
where
\begin{equation}
\Sigma=\;S\;+\sum_{i,m}\;\delta_{i}\;c_{im}\;\log(\eta(\frac{Ti}{l_{im}}))
\label{teli}
\end{equation}
and
\begin{equation}
\Omega(\Sigma)=\;\sum_{a}\;h_a\;e^{\frac{24 \pi^2 \Sigma}{b_a}}
\label{omesi}
\end{equation}
with
\begin{equation}
\delta_i=\frac{\delta^i_{GS}}{8\pi^2}
\label{green}
\end{equation}
and
\begin{equation}
F(T_i)=\;\prod_{i,m}(\eta(\frac{T_i}{l_{im}}))^{-c_{im}}\;
\prod_{j}H_j(T_j)
\label{etafu}
\end{equation}
In what follows, we shall treat $\Sigma$ as a parameter to be
chosen so that $\rm{Re} S$ is a approximately 2, and we shall
also treat
\begin{equation}
\rho\equiv\;\frac{\frac{d\Omega}{d\Sigma}}{\Omega}
\label{param}
\end{equation}
as a free parameter. The parameter $\rho$ is related to the
dilaton auxiliary field $F_S$ by
\begin{equation}
\rho=\;\frac{1-F_S}{y}
\label{aux}
\end{equation}
with
\begin{equation}
y=\;S+\bar{S}-\sum_{i}\;\delta_i\;\log(T_i+\bar{T}_i)
\label{ypar}
\end{equation}

If stabilization of the dilaton expectation value involves
stringy non-perturbative corrections \cite{BANK,casa} to the 
dilaton K$\rm{\ddot{a}}$hler potential then we write the dilaton
and moduli dependent part of the  K$\rm{\ddot{a}}$hler potential as
\begin{equation}
K=\;-\sum_{i}\log(T_i+\bar{T}_i)\;+\;P(y)
\label{pym}
\end{equation}
where $P(y)$ is  a function to be determined by non-perturbative
string effects. In that case, we shall treat $\frac{dP}{dy}$
and $\frac{d^2P}{dy^2}$, which we shall see occur in the effective
potential and soft supersymmetry-breaking terms as free parameters. 
The form of the effective potential that encompass both 
possibilities is
\begin{eqnarray}
V_{eff}&=&\;|W_{np}|^2\;e^{P(y)}\;\prod_{i}(T_i+\bar{T}_i)^{-1} \nonumber \\
&\times&\Biggl\{-3\;+\;\Bigl|\frac{dP}{dy}+\frac{\partial{\log W_{np}}}{
\partial{S}}\Bigr|^2\;(\frac{d^2P}{dy^2})^{-1} \nonumber \\
&+&\sum_{i}(1+\delta_i\frac{dP}{dy})^{-1}\Bigl|(T_i+\bar{T}_i)
\frac{\partial{\log W_{np}}}{\partial{T_i}}-1+\delta_i\frac{
\partial{\log W_{np}}}{\partial{S}}\Bigr|^2\Biggr\}
\label{gepo}
\end{eqnarray}
or, more explicitly,
\begin{eqnarray}
V_{eff}&=&\;|\Omega(\Sigma)|^2\;|F(T_i)|^2 \prod_{j}(T_j+\bar
{T}_j)^{-1}\;e^{P(y)}
\nonumber \\
&\times&\Biggl\{-3+\;\Bigl|\frac{dP}{dy}+\rho\Bigr|^2(\frac{
d^2P}{dy^2})^{-1} \nonumber \\
&+&\sum_{i}(1+\delta_i\frac{dP}{dy})^{-1}(T_i+\bar{T}_i)^2
\Bigl|(\delta_i \rho-1) \hat{G}^i+\frac{d\log H_i}{dT_i}\Bigr|^2
\Biggr\}
\label{telpy}
\end{eqnarray}
where
\begin{equation}
\hat{G}^i=\;(T_i+\bar{T}_i)^{-1}+\;\sum_{m}\;c_{im}\frac{
\partial_{T_i}{\eta\Biggl(\frac{T_i}{l_{im}}\Biggr)}}{\eta
\Biggl(\frac{T_i}{l_{im}}\Biggr)}
\label{einse}
\end{equation}
and in the special case of a single condensate, $\rho$ has the
value $24\pi^2/b$.

The soft supersymmetry breaking terms may be calculated by
standard methods. (See, for example, refs. \cite{ispa}
and \cite{BEATRIZ}, from which the earlier literature can be
traced.) The gaugino masses are given by
\begin{eqnarray}
M_a &=&\;m_{3/2}({\rm{Re}}\;f_a)^{-1}\times\Biggl\{\frac{\partial{
\bar{f_a}}}{\partial{\bar{S}}}(\frac{d^2P}{dy^2})^{-1}\;
(\frac{dP}{dy}+\rho) \nonumber \\
&+&\sum_{i}\frac{(b^{'i}_a-\delta_{GS}^{i})}{8\pi^2}
\frac{y}{(y-\delta_i)}(\rho\delta_i-1)^{-1}(T_i+\bar{T}_i)^2
\Bigl|(\rho \delta_i-1)\hat{G}^i+\frac{d\log H_i}{dT_i}\Bigr|^2\Biggr\}
\label{gluino}
\end{eqnarray}
where $b_{a}^{'i}$ is the usual coefficient occurring in the
string loop threshold corrections to the gauge coupling
constants \cite{LOUIS,ispa} and $f_a$ is the gauge kinetic
function. Provided the dilaton expectation value $S$ and its
auxiliary field in (\ref{gluino}) are real there are no $CP$ violating
phases in the gaugino masses. The soft supersymmetry breaking
$A$ terms are given by
\begin{eqnarray}
m_{3/2}^{-1}A_{\alpha\beta\gamma}&=&\;(\frac{d^2P}{dy^{2}})^{-1}\;
(\frac{dP}{dy}+\bar{\rho})\frac{dP}{dy} \nonumber \\
&-&\sum_{j}(1+\delta_j \frac{dP}{dy})^{-1}(T_j+\bar{T}_j)(
(\delta_j \bar{\rho}-1)\bar{\hat{G}}^j+\frac{d\log \bar{H}_j}{d\bar{T}_j})
\nonumber \\
&\times&(1+n_{\alpha}^{j}+n_{\beta}^{j}+n_{\gamma}^{j}-
(T_j+\bar{T}_j)\frac{\partial{\log h_{\alpha\beta\gamma}}}{\partial{T}_j})
\label{david}
\end{eqnarray}
where the superpotential term for the Yukawa coupling of 
$\phi_{\alpha}$,$\phi_{\beta}$ and $\phi_{\gamma}$ is
$h_{\alpha\beta\gamma}\;\phi_{\alpha} \phi_{\beta}
\phi_{\gamma}$, the modular weights of these states
associated with modular transformations on the moduli
$T_i$ are $n_{\alpha}^{i},n_{\beta}^{i}$ and $n_{\gamma}^{i}$, and
for notational convenience we have included any $U$ moduli as additional
$T$ moduli. The usual rescaling by a factor $e^{K/2}\frac{\bar{W}_{np}}
{|W_{np}|}$ required to go from the supergravity theory derived
from the orbifold compactification of the string theory to the
spontaneously broken globally supersymmetric theory has been
carried out together with normalization of the scalar fields. (See,
for example, ref. \cite{Iba:Spain}.) Throughout we shall assume that
there is no $CP$ violating phase associated with the dilaton
auxiliary field $F_S$ and we shall assume that $\rho$ is a real
parameter.

The expression for the soft supersymmetry breaking $B$ term depends
on the mechanism adopted for generating the $\mu$ term for the
Higgs scalars $H_1$ and $H_2$, with corresponding superfields
$\phi_1$ and $\phi_2$. If we suppose that the $\mu$ term is 
generated non-perturbatively as an explicit superpotential
term $\mu_W\;\phi_1\;\phi_2$ then the $B$ term, which, in this
case we denote by $B_W$ is given by
\begin{eqnarray}
m_{3/2}^{-1}\;B_W&=&\;-1+(\frac{d^2 P}{dy^2})^{-1}\;(\frac{dP}{dy}+
\bar{\rho})(\frac{dP}{dy}+\frac{\partial{\log{\mu_W}}}{\partial{S}})
\nonumber \\
&-&\sum_{j}(1+\delta_j\frac{dP}{dy})^{-1}\;(T_j+\bar{T}_j) \Bigl((
\delta_j\bar{\rho}-1)\bar{\hat{G}}^j+\frac{d\log \bar{H}_j}{d\bar{T}_j}
\Bigr) \nonumber \\
&\times& \Bigl(1+n_1^j+n_2^j-(T_j+\bar{T}_j)\frac{\partial{\log\mu_W}}{
\partial{T_j}}-\delta_j\frac{\partial{\log\mu_W}}{\partial{S}}\Bigr)
\label{bbb}
\end{eqnarray}
where $n_1^j$ and $n_2^j$ are the modular weights of the Higgs
scalar superfields $\phi_1$ and $\phi_2$ and again the appropriate
rescaling of the Lagrangian has been carried out.

On the other hand, if the $\mu$ term is generated by a term of the
form $Z\phi_1\phi_2+(h.c.)$ in the K$\rm{\ddot{a}}$hler potential
mixing the Higgs superfields \cite{anto}, then the tree level form of
$Z$ is
\begin{equation}
Z=(T_3+\bar{T}_3)^{-1}\;(U_3+\bar{U}_3)^{-1}
\label{tez}
\end{equation}
It is assumed that the third complex plane is the $Z_2$ plane with 
whose moduli,
$T_3$ and $U_3$, the untwisted matter fields $\phi_1$ and $\phi_2$
are associated, for this mechanism. Before rescaling the Lagrangian
by $W_{np}/|W_{np}|$ the $B$ term, which in this case we
denote by $B_Z$, is given by
\begin{eqnarray}
-m_{3/2}^{-1}\mu_{Z}^{eff}\;B_{Z}&=&\;W_{np}\;Z\times\Biggl\{2+
\;(T_3+\bar{T}_3)\Bigl(\frac{d\log H_3}{dT_3}-\hat{G}^3(T_3,\bar{T}_3)
+\rm{h.c}\Bigr) \nonumber \\
&+& (U_3+\bar{U}_3)\Bigl(\frac{d\log H_{3}^{'}}{dU_3}-\hat{G}^3(
U_3,\bar{U}_3)+\rm{h.c.}\Bigr) \nonumber \\
&+&\Bigl[(T_3+\bar{T}_3)(U_3+\bar{U}_3)\Bigl(\frac{d\log{\bar{
H}_3}}{d\bar{T}_3}-\bar{\hat{G}}^{3}(T_3,\bar{T}_3)\Bigr)
\Bigl(\frac{d\log{H_3^{'}}}{dU_3}-\hat{G}^3(U_3,\bar{U}_3)\Bigr)+
\rm{h.c}\Bigr]\Biggr\} \nonumber \\
&+&W_{np}\;Z\;\times\;\Biggl\{-3+\;\Bigl|\frac{dP}{dy}+\rho\Bigr|^2\;
(\frac{d^2P}{dy^2})^{-1} \nonumber \\
&-& \sum_{i}\frac{(T_i+\bar{T}_i)^2}{1+\delta_i\frac{dP}{dy}}\;
\Bigl|(\delta_i\rho-1)\hat{G}^i+\frac{d\log H_i}{dT_i}\Bigr|^2
\Biggr\}
\label{deyb}
\end{eqnarray}
where 
\begin{eqnarray}
\mu_Z^{eff}&=&\;|W_{np}|Z\Bigl(1+(T_3+\bar{T}_3)(
\frac{d\log H_3}{dT_3}-\hat{G}^3(T_3,\bar{T}_3)\Bigr) \nonumber \\
&+&(U_3+\bar{U}_3)(\frac{d\log H_3^{'}}{dU_3}-\hat{G}^3(U_3,
\bar{U}_3))\Bigr)
\label{mmm}
\end{eqnarray}
In (\ref{deyb}) and (\ref{mmm}), the factors $H_i$ in (\ref{etafu})
associated with the moduli $T_3$ and $U_3$ have been denoted
by $H_3(T_3)$ and $H_{3}^{'}(U_3)$, respectively. We have also
used the fact that the Green-Schwarz parameters $\delta_i$ 
associated with $T_3$ and $U_3$ are zero (at least for a pure
gauge hidden sector.)The term $Z\phi_1\phi_2+h.c.$ in the K$\rm{\ddot{a}}$hler potential
, with $Z$ given in (\ref{tez}), is merely the first of an infinite series arising from 
the expansion of 
$\log[1+\frac{(\phi_1+\bar{\phi}_2)(\phi_2+
\bar{\phi}_1)}{(T_1+\bar{T}_2)(U_1+\bar{U}_2)}]$
which {\it in total} is modular invariant \cite{anto}. Because the third 
plane is a 
$Z_2$ plane, the $U$-modulus is {\it not} inert under a $T$-modular transformation:
$$U_3 \rightarrow U_3+\frac{\phi_1\phi_2}{icT_3+d}$$
Thus although the whole series is modular invariant, it is not modular 
invariant term by term. It then follows that when we calculate the 
associated soft supersymmetry-breaking $B$-term in the effective potential it 
too is modular invariant, but not term by term. Thus
the best we can do at this juncture is to calculate the $CP$-violating
phase of $B_Z$ at a series of minima of $V_{eff}$ related by modular transformations,
thereby estimating the scale of $CP$-violation to be expected in the calculation of 
observable, and therefore modular invariant, quantities.
In section 5, we shall study the simple model of a single overall
modulus $T$ with
\begin{equation}
T_1=T_2=T_3=T
\label{over}
\end{equation}
and $T_1,T_2$ and $T_3$ on the same footing in the 
non-perturbative superpotential $W_{np}$ and the Yukawa couplings
$h_{\alpha\beta\gamma}$. In addition, we shall assume an unbroken
$PSL(2,Z)$ modular symmetry group for $T$. In that case,
we make the replacements in the above formulae
\begin{equation}
\frac{\partial}{\partial{T_i}}\rightarrow\frac{1}{3}\frac{
\partial}{\partial{T}},\;\delta_i\rightarrow\frac{\tilde{\delta}_{GS}}{
3},\;n_{\alpha}^i\rightarrow n_{\alpha},\;\hat{G}^i\rightarrow \hat{G}
\label{star}
\end{equation}
where
\begin{equation}
\hat{G}(T,\bar{T})=\;(T+\bar{T})^{-1}+\;2 \eta^{-1}\frac{d\eta}{dT}
\end{equation}

If there are non-trivial factors $H_i(T_i)$ in (\ref{etafu})
we also make the replacement
\begin{equation}
\frac{dH_i}{dT_i}\rightarrow \frac{1}{3}\;\frac{dH}{dT}
\label{cusp}
\end{equation}
The assumption of the K$\rm{\ddot{a}}$hler potential mixing
mechanism for $\mu$ is somewhat unnatural in this simple model
because this mechanism requires the presence of a $Z_2$ plane
for the orbifold with an associated $U$ modulus as well as an
associated $T$ modulus. If we wish to employ a model in which
the supersymmetry breaking is dominated by the $T$ moduli we
should then set the auxiliary field for $U_3$ to zero and to take
\begin{equation}
\hat{G}^3(U_3,\bar{U}_3)-\frac{d\log H^{'}_3}{dU_3}=0
\end{equation}
in (\ref{deyb}) and (\ref{mmm}).

\section{The $Z6-IIb$ orbifold with multiple gaugino condensates}

We shall discuss the case of the $Z6-IIb$ orbifold in 
this section and section 4.
This example is rich enough to contain a $T$-modulus ($T_1$) with 
associated $PSL(2,Z)$ target-space modular symmetry, and a pair of 
$T$- and $U$- moduli ($T_3$ and $U_3$) with the congruence subgroups 
$\Gamma^{0}_{T_3}(3)$ and $\Gamma^{0}_{U_3-2i}(3)$ of the associated 
target space modular symmetry groups, where 
$$\Gamma^{0}(n)=\{\left(\begin{array}{cc}
a\;\;\;b \\
c\;\;\; d\end{array} \right),\;\;ad-bc=1,\;b=0\;{\rm mod}\;n\;a,
b,c,d\in Z \}$$
If we assume that the dilaton is stabilized by a multiple gaugino
condensate scenario the effective potential, is given by the
following expression
\begin{eqnarray}
V_{eff}&=&|\Omega(\Sigma)|^{2}|F(T_i)|^{2}
\prod_{i}(T_i+\bar{T}_i)^{-1}y^{-1}\Biggl\{-3+|1-y\rho|^2  \nonumber \\
&+&\sum_{i}\frac{y}{(y-\delta_i)}(T_i+\bar{T}_i)^2 \Bigl|(1-
\delta_i\rho) \hat{G}^{i}+\frac{d\log H_i}{dT_i}
\Bigr|^2\Biggr\}
\label{racetra}
\end{eqnarray}
where
\begin{equation}
F(T_i)=\prod_{i,m}\Bigl(\eta(\frac{T_i}{l_{im}})\Bigr)^{-c_{im}}
\prod_{j}H_j(T_j)
\end{equation}
and $c_{11}=2,c_{31}=c_{32}=c_{41}=c_{42}=1$ and 
$l_{11}=l_{31}=l_{42}=1,l_{32}=l_{41}=3$. 

In the case that the $j$ function is not present in the 
non-perturbative superpotential, one minimizes (\ref{racetra}) 
with respect to the moduli $T_i$ by omitting all the $H_i$ factors.
In this case, numerical minimization gives us the following results.
Let us start with the $T_1$ modulus. For $0 <  \rho \leq 0.4$
the minimum is at a real value of $T_1$ (see Fig.2 and Fig.3) which
approaches 1 as $\rho$ approaches 0.42. For $0.42\leq \rho \leq 0.75$,
$T_1$ remains at the fixed point at $T_1=1$, and for $\rho\geq 0.75$ the
minimum is at the other fixed point at $T_1=e^{\frac{i\pi}{6}}$. (There are
of course also minima at points obtained from these minima by 
modular transformations.
\footnote{
This is a non-trivial check
of the correct modular invariance properties of $V_{eff}$.}
) The other two moduli $T_3,U_3$ behave as follows.
For $0\leq \rho \leq 0.75$ the minima along the $U_{3}$ and $T_3$ directions
in the moduli field space are at the points $\sqrt{3}+i(2+3m)$,
$\sqrt{3}+i(3n)$ respectively, with $n,m\in Z$. The latter points are
zeros of the $\hat{G}^{U_3}$ and $\hat{G}^{T_3}$ functions 
(See Fig.6). (Again there are also minima obtained from 
these minima by modular transformations
\footnote{For example, numerical minimization for
$\rho=0.55$ yields ${\rm T}_{3}|_{min}=\sqrt{3}$, but
also ${\rm T}_{3}|_{min}=0.4330128-0.75 i$. The 
latter point is related to the former through the
modular transformation ${\rm T}_{3} \rightarrow \frac{T_{3}}{i T_{3}+1}$ 
$\in \Gamma^{0}(3)$} .) For $\rho\geq 0.75$ the minima
for both moduli are at their fixed points. More specifically for
$T_3$ the minimum is at $T_3|_{min}=\sqrt{3}/2+ (1.5 +3 n) i$ and for 
$U_3$ at $U_3|_{min}=\sqrt{3}/2+(3.5 +3 m) i$. See also figure 2.
Consequently in the region $0\leq\rho\leq 0.75$ we have anisotropic
solutions while for $\rho\geq 0.75$ we have isotropic solutions.


If, as has been discussed above, we allow the possibility that the
absolute modular invariant $j$ appears in the non-perturbative 
superpotential $W_{np}$, we find 
complex solutions on the unit circle.
For instance for $\rho=0.5$, $\delta^{T_1}_{GS}=-10$, $m_1=n_1=1$,
$m_2=n_2=m_3=n_3=0$ we obtain (see Fig.4) the following solution
\begin{eqnarray}
T_3|_{min}&=&1.73205080+3m\;i \nonumber \\
T_1|_{min}&=&0.97097402\pm 0.23918496\;i \nonumber \\
U_3|_{min}&=&1.73205080+(2+3n)\;i  
\label{phase}
\end{eqnarray}

The soft $A$-terms arising from  (\ref{racetra}) are given by Eq.(\ref{david}),
which reduces to 
\begin{eqnarray}
m_{3/2}^{-1}A_{\alpha\beta\gamma}&=&(1-y\bar{\rho}) \nonumber \\
&-&\sum_{j}\Bigl(\frac{y}{y-\delta_j}\Bigr)(T_j+\bar{T}_j)\Bigl(
(\delta_j\bar{\rho}-1)\bar{\hat{G}}^j+\frac{d\log \bar{H}_j}{d
\bar{T}_j}\Bigr) \nonumber \\
&\times&(1+n_{\alpha}^j+n_{\beta}^j+n_{\gamma}^j-\frac{
\partial{\log h_{\alpha\beta\gamma}}}{\partial{T_j}})
\label{trigra}
\end{eqnarray}
in this case.
The $\frac{\partial{\log h_{\alpha\beta\gamma}}}{\partial{T_j}}$ contribution
to the $A$-terms in (\ref{trigra}) is essential for its modular 
invariance, and can make a significant contribution to any $CP$-violating
phase.  
Non trivial Yukawa couplings from twisted sector 
states arise from $\theta\theta\theta^4$ and $\theta^2\theta^2\theta^2$ 
couplings.
We consider both the Yukawa couplings $h(T_1,k=0)$ where
\begin{equation}
h(T_1,k)\sim e^{-\frac{2}{3}\pi k^2 T_1}\Bigl[
\Theta_3(ikT_1,2iT_1)\Theta_3(ikT_1,6iT_1)+
\Theta_2(ikT_1,2iT_1)\Theta_2(ikT_1,6iT_1)
\end{equation} 
which is covariant (invariant) under $T_1\rightarrow T_1+i$, and the 
linear combination of Yukawas
\begin{equation}
h(T_1,k=0)+(\pm \sqrt{3}-1)h(T_1,k=1)
\end{equation}
which is covariant under $T_1\rightarrow \frac{1}{T_1}$ (see the discussion
in $\S 5$.)

Similarly the soft susy-breaking $B_W$-term, arising from 
(\ref{racetra}) is given by (\ref{bbb}), which reduces to
\begin{eqnarray}
m_{3/2}^{-1}B_W&=&-1+(-1+y\bar{\rho})(-1+y\frac{\partial{\log\mu_W}}
{\partial{S}}) \nonumber \\
&-&\sum_{j}(\frac{y}{y-\delta_j}) \;(T_j+\bar{T}_j) \Bigl((
\delta_j\bar{\rho}-1)\bar{\hat{G}}^j+\frac{d\log \bar{H}_j}{d\bar{T}_j}
\Bigr) \nonumber \\&\times& \Bigl(1+n_1^j+n_2^j-(T_j+\bar{T}_j)
\frac{\partial{\log\mu_W}}{
\partial{T_j}}-\delta_j\frac{\partial{\log\mu_W}}{\partial{S}}\Bigr)
\label{phroso}
\end{eqnarray}
in this case, where \cite{anto} $\mu_W$ is given by
\begin{equation}
\mu_{W}\propto W_{np} \frac{\partial{\log \eta(T_3)\eta(\frac{T_3}{3})}}
{\partial{T_3}}\frac{\partial{\log \eta(U_3)\eta(\frac{U_3}{3})}}{
\partial{U_3}}
\label{niadis}
\end{equation}
Although this form of $\mu_W$ has the correct behaviour under $T_3$-
and $U_3$-modular transformation, the resulting expression for 
$B_W$ is not modular invariant; (the reason for this is that exact 
modular invariance requires the use of the one-loop corrected 
K$\rm{\ddot{a}}$hler potential \cite{anto}.) 

The values given above for the moduli $T_i$ at the minimum of $V_{eff}$
, when the Dedekind $\eta$ function is the only modular function 
present in $W_{np}$,
have rather striking consequences for the possible $CP$ violating phases 
in the soft supersymmetry-breaking terms. It
might have been thought {\it a priori} that when $T_i$ is at a fixed
point at $e^{\frac{i\pi}{6}}$ a $CP$ violating phase of order $10^{-1}$ 
might be induced. However, as has been observed earlier \cite{steve}, if
$T_i$ is precisely at a fixed point value, and/or at a zero of 
$\hat{G}(T_i,\bar{T}_i)$, the $CP$ violating phase vanishes identically,
as can be seen from (\ref{gluino})-(\ref{deyb}). To illustrate the 
importance of the latter statement we have plotted the dependence of the
imaginary part of the Eisenstein function 
$\hat{G}^{T_3}$ of the $T_3$ modulus, with respect to the imaginary 
part of the modulus, for fixed real part (see Fig. 5) of the latter.
We note from the graphs that if the moduli are exactly at a fixed 
point,
at the minimum of the 
potential energy,  zero phases occur while 
even a small departure from the imaginary 
part at $Im T_3=1.5$ in our example, can cause an imaginary part for
$\hat{G}^{T_3}$ of order $10^{-2}$ which can be fed to 
the soft-supersymmetry breaking terms as a phase of order $10^{-2}$.
Later, when we discuss the model in which non-perturbative 
corrections to the K$\rm{\ddot{a}}$hler potential are responsible for
the dilaton stabilization, we will see that there is another mechanism 
for suppressing $CP$ phases which is due to the very rapid variation of the 
imaginary part of $\hat{G}^{i}$ with $\rm{Re} T$ as Re $T$ moves away 
from 1 if Im $T$ is held fixed. 

For the 
multiple gaugino condensate scenario in the $Z6-IIb$ orbifold we 
conclude the following: If the Dedekind $\eta$ function is the 
{\it only modular form} appearing in the non-perturbative superpotential
there are no $CP$ violating phases in the soft-supersymmetry breaking 
terms.
 
If, on the other hand, the $j$ function is also involved in $W_{np}$ 
besides the $\eta$ function, then $CP$ violating phases may arise. For 
the Yukawa couplings
we have considered, and for the particular solution presented in 
Eq.(\ref{phase}), we obtain  $CP$ violating phases not greater than
$2\times 10^{-5}$ in the $A$ term. However, the
 $B_W$ soft superymmetry-breaking terms lead to a zero phase for the solution 
presented in (\ref{phase}). Choi 
\cite{choi} has also 
obtained small $CP$- 
violating phases in a class of multiple gaugino condensate models
which possess an approximate Peccei-Quinn symmetry.

Thus overall,  in the multiple gaugino condensate scenario 
in the $Z6-IIb$ orbifold, we conclude that if both the $j$-function 
and the Dedekind  $\eta$ function are present in  $W_{np}$ 
then $CP$-violating 
phases not greater than $10^{-5}$ arise.

\section{The 
$Z6-IIb$ orbifold model with non-perturbative corrections to 
the dilaton K$\rm{\ddot{a}}$hler potential}

In this section we continue the discussion of the $Z6-IIb$ orbifold 
introduced in section 3.
If the dilaton VEV is stabilised by non-perturbative corrections to the
K$\rm{\ddot{a}}$hler potential, 
we minimize the effective potential
with the modular invariant $y$ 
fixed at 4, for different values of the parameters
$\frac{dP}{dy}$ and $\frac{d^2P}{dy^2}$. 
The effective potential in this case is given by the following expression
\begin{eqnarray}
V_{eff}&=&|\Omega(S)|^2 |F(T_i)|^2 \prod_{i}(T_i+\bar{T}_i)^{-1}
e^{P(y)} \nonumber \\
&\times& \Biggl\{-3+\Bigl|\frac{dP}{dy}+\frac{24\pi^2}{b}\Bigr|^2
(\frac{d^2P}{dy^2})^{-1} \nonumber \\
&+&\sum_{i}(1+\delta_i\frac{dP}{dy})^{-1}(T_i+\bar{T}_i)^{2}
\Bigl|(\delta_i \frac{24\pi^2}{b}-1)\hat{G}^i+
\frac{d\log H_i}{dT_i}\Bigr|^2\Biggr\}
\label{arxi}
\end{eqnarray}
with
\begin{equation}
\Omega(S)=h\;e^{24\pi^2 S/b}
\end{equation}
and 
\begin{equation}
F(T_i)=\prod_{i,m}\Bigl(\eta\Bigl(\frac{T_i}{l_{im}}\Bigr)\Bigr)^{
-c_{im}(1-\frac{3\delta^{i}_{GS}}{b})}\prod_{j}H_j(T_j)
\end{equation}
and $c_{11}=2,c_{31}=c_{32}=c_{41}=c_{42}=1$ and $l_{11}=
l_{31}=l_{42}=1,l_{32}=l_{41}=3$.

In this case we obtain the
following results. First, if the non-perturbative superpotential $W_{np}$
includes only the Dedekind eta function, the values of the $T_1$ modulus
for a wide range of the parameters are either real or at a fixed point of $PSL(2,Z)$.
On the other hand, the values at
the minimum of the effective potential for the $T_3$ and $U_3$ moduli
are either at the points $\sqrt{3}+ i 3p$, $\sqrt{3}+i (2+3m)$ 
($p,m$ are integers) respectively,
or they are at the fixed points of $\Gamma^{(0)}(3)$. 


Second, if we allow for the possibility that the $j$ function appears in the
form of the non-perturbative superpotential as well as the $\eta$ function,
we obtain solutions at complex values of the moduli on the {\it unit
circle}. For instance for $\frac{dP}{dy}=-1/4,\;\frac{d^2P}{dy^2}=0.6$,
$m_1=n_1=1, m_i=n_i=0$ for $i=2,3$ and $\tilde{\delta}_{GS}^{T_1}=-10$ (see fig.8) the 
following solution occurs 
\begin{eqnarray}
T_1|_{min}&=&0.97090183\pm 0.23947783\;\;i  \nonumber \\
T_3|_{min}&=&0.8660254+1.49999999\;\;i \nonumber \\
U_3|_{min}&=&0.8660254+3.5000000\;\;i
\label{Z6}
\end{eqnarray}
In this case, the minimum on the unit circle for the $T_1$ modulus
is at a  zero of the quantity 
\begin{equation}
(\delta_i\frac{24\pi^2}{b}-1)\hat{G}^i+\frac{d\log H_i}{dT_i}
\label{zerom}
\end{equation}
 appearing in (\ref{arxi}). This can be demonstrated analytically and
agrees with the numerical analysis. In consequence,
 as we will discuss below, the $CP$-violating phases are zero 
for minima of the $V_{eff}$ which are zeros of (\ref{zerom}).  

The soft terms arising from (\ref{arxi}) are given by

\begin{eqnarray}
m_{3/2}^{-1}A_{\alpha\beta\gamma}&=&\Bigl(\frac{d^2 P}{dy^{2}}\Bigr)^{-1}
\Bigl(\frac{dP}{dy}+\frac{24\pi^2}{b}\Bigr)\frac{dP}{dy} \nonumber \\
&-&\sum_{j}\Bigl(1+\delta_j\frac{dP}{dy}\Bigr)^{-1}(T_j+\bar{T}_j)
\Bigl((\frac{3\delta^{j}_{GS}}{b}-1){\bar{\hat{G}}}^{j}+\frac{d\log \bar{H}_j
}{d\bar{T}_j}\Bigr) \nonumber \\
&\times&\Bigl(1+n_{\alpha}^{j}+n_{\beta}^{j}+n_{\gamma}^{j}-
(T_j+\bar{T}_j)\frac{\partial{\log h_{\alpha\beta\gamma}}}{
\partial{T_j}}\Bigr)
\label{tigri}
\end{eqnarray}

If the
Dedekind $\eta$ function is the only modular function that appears 
in $W_{np}$, we find that the $CP$-violating phases are either zero or much smaller 
than $10^{-3}$.

If the $j$-function is present in $W_{np}$, together with the Dedekind $\eta$ function,
and we use solutions such as (\ref{Z6}), then the $CP$-violating phases are zero since,
as noted above, these complex minima are zeros of(\ref{zerom}). However, for different 
values of $\frac{dP}{dy}$ and $\frac{d^2P}{dy^2}$ there are other minima where 
significant 
$CP$-violating phases arise. For example, for 
$\frac{dP}{dy}=-2.45,\frac{d^2P}{dy^2}=-0.45,m(T_1)=n(T_1)=1,
m(T_3)=n(T_3)=n(U_3)=m(U_3)=0$, $\delta_{GS}^{T_1}=-20$ we obtain 
the following 
solution (see Fig.7)
\begin{eqnarray}
T_3|_{min}=0.31054252+\;2.999998\;i \nonumber \\
T_1|_{min}=0.97084106\pm 0.23972408\;i \nonumber \\
U_{3}|_{min}=0.07838707+1.84680549\;i
\label{photia}
\end{eqnarray}
This solution leads to a $CP$-violating phase $\phi(B_W)$ of order 
$10^{-1}$. On the other hand $\phi(A)$ is of order $10^{-2}$.  

\section{The single overall modulus case}

In this case the relevant formulae for the effective potential and soft 
supersymmetry-breaking terms  are obtained 
by making the substitutions (\ref{star})-(\ref{cusp}) (See also 
\cite{BKL}. 

The $\frac{\partial{\log h_{\alpha\beta\gamma}}}{\partial{T}}$  contribution
to 
the $A$-terms deriving from (\ref{david}) is essential for its modular invariance 
and can make a significant contribution to
any $CP$ violating phase. For illustrative purposes we have taken
$h_{\alpha\beta\gamma}$ to be of the form encountered \cite{Chun}
 when each of the states $\phi_{\alpha},\phi_{\beta}$ and 
$\phi_{\gamma}$ is in the particular twisted sector of the
$Z_3\times Z_6$ orbifold with the same twisted boundary conditions
 as the twisted sector of the $Z_3$ orbifold. This is an appropriate choice
because the $Z_3\times Z_6$ orbifold has three $N=2$ moduli, $T_i$, $
i=1,2,3$, so that the model of a single overall modulus $T=T_1=T_2=T_3$ is
consistent. In this case, if we arrange $h_{\alpha\beta\gamma}$ to be covariant
under the $T\rightarrow \frac{1}{T}$ modular transformation, it is a product
of 3 factors, one for each complex plane, of the form
\begin{equation}
h(T_i,k_i=0)+(\pm \sqrt{3}-1)h(T_i,k_i=1)
\end{equation}
where
\begin{equation}
h(T_i,k_i)\sim e^{-\frac{2}{3}\pi k_i^2 T_i}\Bigl[
\Theta_3(ik_iT_i,2iT_i)\Theta_3(ik_iT_i,6iT_i)+
\Theta_2(ik_iT_i,2iT_i)\Theta_2(ik_iT_i,6iT_i)\Bigr]
\end{equation}
Each of the modular weights $n_{\alpha},n_{\beta}$ and $n_{\gamma}$ has
the value -2. Yukawa couplings associated with untwisted 
matter fields have modular weight 0 and are generically $T$-independent constants.

In the case of multiple gaugino condensates with perturbative
K$\rm{\ddot{a}}$hler potential, minimization of the
effective potential at fixed $\Sigma$ for different
real values of the parameter $\rho$ with Re S taken to
be about 2 leads to the following conlusions.
It can be seen analytically that the fixed points 
of $PSL(2,Z)$ at $T=1$ and $T=e^{\frac{i\pi}{6}}$, at
which $\hat{G}(T,\bar{T})$ is zero, are always
extrema  (even for $\tilde{\delta}_{GS} \neq 0$.)
For $0.1\leq\rho\leq 0.4$, the minimum is at a real value
of $T$ which approaches 1 as $\rho$ approaches 0.42.
For $0.42 \leq \rho \leq 0.75$, $T$ remains at the fixed 
point at $T=1$, and for $\rho\geq 0.8$ the minimum is at
the other fixed point at $T=e^{\frac{i\pi}{6}}$
(See Fig.9). (There
are of course also minima at points obtained from
these minima by modular transformation.) This resembles
what happens for a single condensate but treating
the dilaton auxiliary field $F_S$ as a free parameter
to simulate dynamics stabilizing the dilaton expectation
value \cite{ferrara}.

Gaugino condensate models (with perturbative 
K$\rm{\ddot{a}}$hler potential) in general have 
negative vacuum energy at the minimum. However, as other
authors have emphasized \cite{ferrara}, the solution 
to the vanishing cosmological constant problem is probably
in the realm of quantum gravity and as the present type
of discussion treats gravity classically we need not
necessarily impose vanishing vacuum energy as a constraint
on the theory. On the other hand, if we do arrange for
zero vacuum energy by introducing an extra matter field
which does not mix with the dilaton and moduli
fields \cite{Iba:Spain} then the effect in the minimization
of the effective potential with respect to $T$ is that the
factor premultiplying the bracket in (\ref{telpy}) is not
to be differentiated.
Then, for $0.25\leq \rho \leq 2.15$ minima occur
at the fixed points  at $T=1$ and $T=e^{\frac{i \pi}{6}}$.
For $\rho \geq 2.2$ there is a single real minimum.

In the case of a single gaugino condensate, but with
the dilaton expectation value being stabilized by
stringy non-perturbative corrections to the dilaton
K$\rm{\ddot{a}}$hler potential, minimization of the
effective potential with $y$ fixed at 4 for different
values of the parameters $\frac{dP}{dy}$ and 
$\frac{d^2P}{dy^2}$ leads instead to the following
outcome. For a wide range of choices of these parameters
$T$ is either at a fixed point value or it is real.
However, for $\frac{dP}{dy}=-11/4$ and 
$\frac{d^2P}{dy^2}=-1.3$, one obtains
\begin{equation}
T|_{min}=5.234339+0.0009575 i
\label{parad}
\end{equation}
and the potential is more flat.

The consequences of the values of the modulus $T$ at the
minimum for possible $CP$ violating phases in the soft
supersymmetry breaking terms are 
similar to those of the $Z6-IIb$ orbifold
Thus, for the case where the dilaton is stabilized
by a multiple gaugino condensate with perturbative 
K$\rm{\ddot{a}}$hler potential, there are no
$CP$ violating phases in the soft supersymmetry breaking
terms. In the case of a single gaugino condensate
with the dilaton stabilized by non-perturbative
corrections to the dilaton K$\rm{\ddot{a}}$hler potential,
the conclusion is the same for a wide range of values
of $\frac{dP}{dy}$ and $\frac{d^2P}{dy^2}$.
However, in this case, it is possible for the minimum
for $T$ to be at a complex value away from the fixed
point as, for example, in (\ref{parad}). At first sight,
there might then be a $CP$ violating phase of order
$10^{-3}$.
However, the $CP$ violating phases are far smaller than this
(of order $10^{-15}$ for $T$ as in (\ref{parad}).)
The reason for this is the very rapid variation of the
imaginary part of $\hat{G}(T,\bar{T})$ with $Re T$ 
as Re $T$ moves away from 1 if Im $T$ is held fixed
(see fig. 10). The imaginary part of 
$\hat{G}(T,\bar{T})$ varies by 11  orders of
magnitude as Re $T$ goes from $\frac{\sqrt{3}}{2}$
to 5.0. The possibility of suppressing $CP$ violating
phase in this way has been suggested earlier
\cite{steve} in the context of orbifold models
with broken $PSL(2,Z)$ modular symmetries.

If we allow the possibility
that the non-perturbative superpotential $W_{np}$ involves
the absolute modular invariant $j(T)$ as well as the 
Dedekind eta function \cite{miriam} then the situation
is very different. Then, $W_{np}$  contains an extra factor
$H(T)$ where the most general form of $H(T)$ to avoid
singularities inside the fundamental domain \cite{miriam} is
\begin{equation}
H(T)=(j-1728)^{m/2}j^{n/3} P(j)
\label{soti}
\end{equation}
where $m$ andf $n$ are integers and $P(j)$ is a polynomial
in $j$. It is then possible, for some choices
of $H$ to obtain (complex) minima of the effective potential
for $T$ that lead to $CP$ violating phases in the
soft supersymmetry breaking terms of order $10^{-4}-10^{-1}$.
Let us start with the case of stabilizing the dilaton
by multiple gaugino condensate, then for $P(j)=1$ and $m=n=1$,
$\tilde{\delta}_{GS}=-\frac{30}{8 \pi^2}, \; \rho=0.45$,
we find 
that the minimum is on the unit circle at
\begin{equation}
T|_{min}=0.971353713\pm 0.2376383050 i
\end{equation}
For the Yukawa couplings which 
we have considered (see below), this  leads to a $CP$ violating phase
not greater than $10^{-4}$. 
Also for $P(j)=1$ and $m=n=1$,
$\tilde{\delta}_{GS}=-\frac{50} {8 \pi^2}, \; \rho=0.26$
the minimum is 
also on the unit circle at
\begin{equation}
T|_{min}=0.971352323\pm 0.237643985 i
\end{equation}
which again leads to $CP$ violating phase not greater
than $10^{-4}$ in the $A$ term.
On the other hand, if we assume that the dilaton is stabilized by 
non-perturbative corrections to the K$\rm{\ddot a}$hler potential,
then it is possible to find minima of the effective potential at
complex values of the $T-$ modulus not only on the boundary of
the fundamental domain {\it but also inside the fundamental domain}.
In this case larger $CP$ violating phases arise.

Let us start with the solutions on the unit circle.
For $m=n=1$ we can obtain for a wide range of parameters for 
$\frac{dP}{dy}$ and $\frac{d^2P}{dy^2}$ solutions on the unit 
circle. For instance a representative solution is:
\begin{equation}
T|_{min}=0.971143367\pm 0.238496458\;i
\end{equation}
Where $\frac{dP}{dy}=2.45$ and $\frac{d^{2}P}{dy^2}=0.45$ is our 
choice for our non-perturbative parameters. 
The $CP$-violating phases in this case are zero because as we 
argued in $\S 4$ these solutions correspond to zeros of
Eq.(\ref{zerom}). We can also obtain
local minima on the unit circle for the following choice of 
$(m,n)=(1,3)$. Let us a give an example:
\begin{equation}
T|_{min}=0.98650708\pm 0.163718557\;\;i
\end{equation}
for $\frac{dP}{dy}=-1.15$ and $\frac{d^{2}P}{dy^2}=0.45$
In this case however and in general for $n\geq 2$ the $T=e^{\frac{i\pi}{6}}$ 
is 
a minimum
with zero cosmological constant.

Let us now
describe , 
solutions not on the 
boundary of the fundamental 
domain but inside where it is possible for some values of the parameters
$\frac{dP}{dy}$ and $\frac{d^2P}{dy^2}$ to obtain phases that exceed
the current experimental limit. As a result we can constrain our
non-perturbative parameter space.
For $\frac{dP}{dy}=-1.5$ and 
$\frac{d^2P}{dy^2}=-0.2,\;m=1,n=3,\delta_{GS}=-30$ we obtain
the following global minima :
\begin{equation}
T|_{min}=1.01196232+ 0.16800043\;i
\label{fird}
\end{equation}
and its $T-dual$ under the generator $T\rightarrow \frac{1}{T}$,
\begin{equation}
T|_{min}=0.96167455-0.15965193\;i
\label{secd}
\end{equation}
Of course as we said before we can obtain minima connected to the
above by  modular transformations as well. \footnote{
Again the appearance of minima connected by modular 
transformations is a non-trivial check of the correct 
implementation of modular invariance properties in our 
expression for $V_{eff}$.}

Both minima lead to a phase $\phi(A)$  of order $10^{-2}$.
The foregoing results need a little amplification. 
For minima connected by $T\rightarrow T+i$ 
the Yukawa $h_{\alpha\beta\gamma}=h(T,k=0)$ leads 
to the same $CP$ violating phases at both minima, while 
for minima connected by $T\rightarrow\frac{1}{T}$ the 
Yukawa $h_{\alpha\beta\gamma}=
h(T,k=0)+(\sqrt{3}-1)h(T,k=1)$, which 
transforms as $h_{\alpha\beta\gamma}(1/T)=Th_{\alpha\beta\gamma}(T)$, also 
leads to the same $CP$ violating phases
at both minima. Both are of order $10^{-2}$.
However, since there is no linear combination of 
Yukawas which has modular weight 1
with respect to {\it all} modular transformations, we cannot do
better than characterize the scale of the CP violating  phases in this
way. It is important to note that since $V_{eff}$ is modular invariant,
the calculation of the electric dipole moment of the neutron,
for example, will necessarily yield a modular invariant
result, presumably 
with magnitude characteristic of the order  
$10^{-2}$ scale of the $CP$ violating phase of $A_{\alpha\beta\gamma}$.
This calculation will necessarily entail contributions from more than
one $A$ term. 
For the $B_W$ soft term we similarly obtain a phase at both of the minima 
given in Eq.(\ref{fird})-(\ref{secd}) of order $10^{-3}$.

Similarly, for $\frac{dP}{dy}=-1.4,\frac{d^2P}{dy^2}=-0.1,\;m=1,n=3,
\delta_{GS}=-30$ (see Fig. 11) we obtain
\begin{equation}
T|_{min}=0.79314323+ 0.11307387\;i
\label{t1}
\end{equation}
its $T-dual$ under $T\rightarrow \frac{1}{T}$
\begin{equation}
T|_{min}=1.23569142-0.17616545\;i
\label{t2}
\end{equation}
as well as the $T-dual$ of the later under $T\rightarrow T+i$
\begin{equation}
T|_{min}=1.23569142+0.82383457\;i
\label{t3}
\end{equation}
At the above three points of the moduli space we find a phase
$\phi(A)$ of order $10^{-3}-10^{-2}$. On the other hand the 
$CP$ violating phase $\phi(B_W)$ 
and $\phi(B_Z)$ are of order $10^{-2}-10^{-1}$ at the 
three points of the moduli space.

For $m=n=1$,$\delta_{GS}=-30$,$\frac{dP}{dy}=-2.45,
\frac{d^2P}{dy^2}=-0.45$ we get
the following representative solutions (see Fig. 12):
\begin{equation}
T|_{min}=0.6649428+0.2387706\;i
\end{equation}
its $T-dual$ under $T\rightarrow \frac{1}{T}$ 
\begin{equation}
T|_{min}=1.332122-0.4783445\;i
\end{equation}T
as well as the $T-dual$ of the later under $T\rightarrow T+i$
\begin{equation}
T|_{min}=1.332122+0.5216555\;i
\end{equation}
In this case at the 3 points of the moduli space we obtain a phase
$\phi(A)$ of order $10^{-1}$. On the other hand $\phi(B_W)$ is 
of order $10^{-2}-10^{-1}$ in the above points of the moduli space.
Also $\phi(B_Z)$ is of order $10^{-3}-10^{-1}$.

Let us summarize the dependence of the minima on the integers 
$m,n$.
For $m=0,n=3$, $T=e^{\frac{i\pi}{6}}$ is a minimum with $V=0$. The
other fixed point $T=1$ is a minimum with either positive or negative
energy dependening on the size and sign of the stringy non-perturbative
parameters $\frac{dP}{dy},\frac{d^2P}{dy^2}$. We also get real minima 
of $O(1)$.
For $m=n=2$ both fixed points are mimima with zero vacuum energy.
We also have solutions on the unit circle with negative or positive 
energy depending on the stringy parameters.
For $m=0,n=2$ $T=e^{\frac{i\pi}{6}}$ is a minimum with $V=0$.
Again the other fixed point $T=1$ is a minimum with either positive or negative
energy depending on the size and sign of the stringy non-perturbative
parameters $\frac{dP}{dy},\frac{d^2P}{dy^2}$.
For $m=2,n=0$ $T=1$, is a zero $V$ minimum. We also get solutions on 
the unit circle.
For $m=1,n=2$, $T=e^{\frac{i\pi}{6}}$ is a zero energy minimum.
Solutions on the unit circle with are also obtained.
However, in this case we also get minima 
with negative energy inside the fundamental domain
of the $T$-modulus (see Fig. 1).
For instance, for $\frac{dP}{dy}=-1.5,\frac{d^2P}{dy^2}=-0.15$ 
we get 
\begin{eqnarray}
T|_{min}&=&1.2819602+0.2350069\;i \nonumber \\
T|_{min}(1/T)&=&0.7546934-0.13834922\;i
\end{eqnarray}
For this particular solution both $A$ and $B$ soft supersymmetry-breaking
terms lead to a phase of order $10^{-1}$ at both points of the 
moduli space.
For $m=1,n=3$ as we saw above we get global minima inside the fundamental
domain, zero energy minima at $T=e^{\frac{i\pi}{6}}$, plus solutions on the
unit circle.
For $m=1,n=1$ solutions inside the fundamental domain plus solutions on the
unit circle.
For $m=3,n=0$, zero energy minima at $T=1$, and solutions on the
boundary of the fundamental domain of the form $T|_{min}=
O(1)+0.5\;i$ together with their $T-$ duals. 

In conclusion, whether the dilaton expectation value is
stabilized by a multiple gaugino condensate or by
stringy corrections to the dilaton 
K$\rm{\ddot{a}}$hler potential, we have found that ,
provided the superpotential does not contain the
absolute modular invariant $j((T)$, $CP$ violating phases
in the soft supersymmetry-breaking
terms are either zero
or much smaller than $10^{-3}$.
Zero phases occur when the minimum
for the modulus $T$ is at a zero of $\hat{G}(T,\bar{T})$ or
is real.
Phases much smaller than $10^{-3}$ occur with $T$ at a complex
value with the real part of $T$ far from its value
at a zero of $\hat{G}(T,\bar{T})$, because of the rapid
variation of the imaginary part of  $\hat{G}(T,\bar{T})$
as  $Re\;T$ varies. However, if we
allow the more general possibility that $W_{np}$
involves $j(T)$
as well as $\eta(T)$, as may arise from orbifold theories
if the theory contains gauge non-singlet states that
are zero at some special values of the moduli
\cite{miriam} or may arise from $F$-theory compactifications
\cite{witten}, then it is possible in some models to obtain
$CP$ violating phases in the soft supersymmetry breaking terms
of order $10^{-4}-10^{-1}$.
The largest phases occur for 
 minima of the potential inside
the fundamental domain of the $PSL(2,Z)$ $T-$ modulus.

\section{Conclusions}
In this work we studied the CP-violating properties 
of soft supersymmetry-breaking terms emerging from target 
space modular invariant effective string supergravities and 
in which supersymmetry
is broken by non-perturbative corrections to the dilaton 
K$\rm{\ddot{a}}$hler potential (stringy-effects) 
and gaugino-condensation (field theoretic non-perturbative effects). We found that the
 $CP$ properties
depend on the 
properties of the modular functions involved
as well as on the mechanism  for stabilizing the dilaton.  The following
remarkable picture emerges. In the case of racetrack models 
(multiple gaugino condensates) with
modular symmetries which are 
subgroups of $PSL(2,Z)$ as well as with the 
full $PSL(2,Z)$ modular symmetry (overall modulus
case), if the Dedekind $\eta$ function is the {\it only modular form} 
appearing in the $W_{np}$ there are no $CP$-violating phases in the
resulting soft supersymmetry-breaking terms. This is the case because
minimization of the effective potential, $V_{eff}$, results in either
real values for the orbifold moduli $T_i$ or complex values at the fixed 
points of the duality group. 
Complex values at the duality group fixed points correspond to zeros of the
Eisenstein functions $\hat{G}^{i}$ appearing in the expressions for the 
soft susy-breaking terms. 
As a result, in both cases the soft supersymmetry-breaking
terms {\it are real}. In the case that the dilaton is stabilized by 
non-perturbative stringy
corrections to the K$\rm{\ddot{a}}$hler potential,
with the $\eta$ function again the only modular function
appearing in $W_{np}$, the resulting $CP$ phases are either zero or much
smaller than of order $10^{-3}$. 

The strong experimental upper bound on the electric dipole moment of the neutron
 translates into bounds on the $CP$-violating phases of the soft supersymmetry-breaking
$A$ and $B$ terms $\phi(A),\phi(B)\leq O(10^{-3})$. The results we have obtained afford
 the possibility of a stringy explanation of these bounds. 
As we emphasized in the introduction, the dipole moment tends
 to be two to three orders of magnitude
too large in generic supersymmetric models.

In the case that the absolute modular invariant function $j$ appears together
with the $\eta$ function in $W_{np}$, and in the case of racetrack 
models,
for a large region of
the parameter space
the $CP$ phases are suppressed below the experimental limit 
even for complex values of  the moduli $T_i$ on the unit circle
at the minimum, and in some regions are close to the current experimental
limit;(we find phases less than about $10^{-4}$). In the case, that the
dilaton is stabilized by non-perturbative stringy 
corrections to the K$\rm{\ddot{a}}$hler potential, then in some regions of
the parameter space larger $CP$-violating phases can arise, of 
order $10^{-3}-10^{-1}$ 
for values of
the moduli {\it inside the fundamental domain} of the modular group.
Our solutions at the minimum of the effective potential at complex values
for the orbifold moduli $T_i$ inside the fundamental domain, provide 
a counterexample to the conjecture of \cite{miriam} that the minima of
the target space duality invariant potential occur only on the boundary
of the fundamental domain of the modular group.
Therefore, in this case we obtain constraints on the non-perturbative 
K$\rm{\ddot{a}}$hler potential parameters from 
resulting phases that exceed the current experimental limits. However,
we must emphasize again that even in the above case, there 
exist  CP-violating 
phases in the soft-supersymmetry breaking 
terms ,
which have the correct order of magnitude for  {\it explaining} \cite{abel}
the severe experimental limits
coming from the electic dipole moment of the neutron and, potentially ,
the tiny observed CP violation \cite{nekaon}.

Our work, motivates the 
investigation \cite{us} of other duality invariant string
theories in which modular functions of higher genus are involved. As
has been shown in \cite{niste}, genus-2 theta functions appearing in the
threshold corrections to the gauge couplings 
which include dependence on the continous Wilson
line moduli , can explain the discrepancy
of the string scale unification with the observed scale coming from 
extrapolation of the LEP results. The values of the moduli in this case 
are close to the ones favoured by duality invariant gaugino condensates. 
Also, the study of $CP$-violation in the 
Kobayashi-Maskawa matrix  in this class of models deserves
investigation \cite{us}. As we saw the Yukawa couplings for twisted matter 
fields transform non-trivially under the modular group and can therefore 
develop complex phases.

\section{Acknowledments}
This research is supported in part by PPARC.

\appendix
\section{Modular functions}

We list here some useful formulae for the modular functions
involved in our work. We first discuss the case of the full modular 
symmetry $PSL(2,Z)$.

A modular form of weight $k$ is a 
holomorphic function $f$,that under a 
generic modular transformation 
\begin{equation}
T\rightarrow \frac{aT-ib}{icT+d},\;\;\;\;a,b,c,d\in Z,\;\;\;\;ad-bc=1
\label{modular}
\end{equation}
transforms as
\begin{equation}
f\Biggl(\frac{aT-ib}{icT+d}\Biggr)=(icT+d)^{k}f(T)
\end{equation}
Typical modular forms are the {\it Eisenstein series} in terms of
which the $\eta(T)$ and $j(T)$ functions can be represented.
Thus an Eisenstein series is a modular form of weight $k$
given by the expression
\begin{equation}
G_k(T)=2\zeta(k)\Bigl(1-\frac{2k}{B_k}\sum_{n\geq 1}
\sigma_{k-1}(n)\;e^{-2\pi  n T}\Bigr)
\label{eisen}
\end{equation}
with $\zeta(k)$ is Riemann's zeta function, $B_k$ are the 
Bernoulli numbers and $\sigma_{p}(n)$ is the sum of the 
$p$-powers of all divisors of $n$.

The absolute discriminant $\Delta$ is a modular form of weight $12$
and is given by
\begin{equation}
\Delta=\frac{675}{256 \pi^{12}}\Bigl(20 G_4^3-49 G_6^2\Bigr)
\label{Delta}
\end{equation}
while the absolute modular invariant $j$ is given by the equation
\begin{equation}
j=\frac{3^6 5^3}{\pi^{12}}\frac{G_4^3}{\Delta}
\end{equation}

By a theorem,
proven in \cite{leshner}, every modular form $f(T)$, regular in the fundamental 
domain, can be written as follows:
\begin{equation}
f(T)=(G_6(T))^s (G_4(T))^{t} (\eta(T))^{2 r -12 s-8 t} P(J)
\end{equation}

For every modular form $f$, of weight $k$, the following derivative
\begin{equation}
D_T f(T)=f^{'}(T)+\frac{k}{2\pi}G_2(T) f(T)
\label{modula}
\end{equation} 
is a modular form of weight $k+2$. This can be proven by using 
the definition of a modular form and the relation of ${G}_2$ with 
the logarithmic derivative of the $\eta$ function.  

\begin{figure}[h]
\input{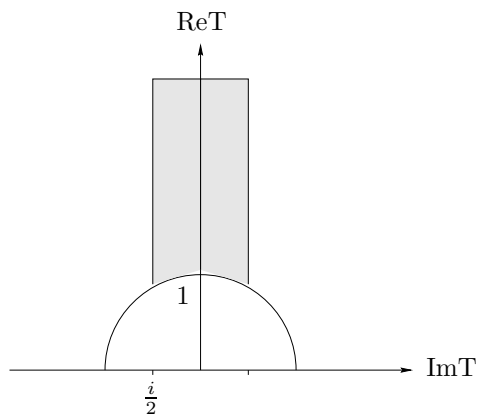}
\caption{Fundamental domain $\Gamma$ of the 
full modular group}
\end{figure}


The congruence subgroup $\Gamma^{0}(3)$ has as generators the 
transformations
$T\rightarrow T+3\;i$, $T\rightarrow \frac{T}{iT+1}$,$I$.


\newpage
\begin{figure}
\epsfxsize=7.25in
\epsfysize=4.5in
\epsfxsize=7.25in
\epsfysize=4.5in
\caption{Behaviour of the real parts of $T_1$ and $T_3$ moduli with
respect to the parameter $\rho$ in multiple gaugino condensate models}
\end{figure}
\newpage
\begin{figure}
\epsfxsize=7.25in
\epsfysize=4.5in
\epsfxsize=7.25in
\epsfysize=4.5in
\caption{Dependence of the imaginary parts of $T_1$ and $T_3$ moduli 
with $\rho$ as in Figure 2.}
\end{figure}
\newpage
\begin{figure}
\epsfxsize=6.0in
\epsfysize=8.0in
\caption{Solutions on the unit circle in the multiple gaugino 
condensate case.}
\end{figure}
\newpage
\begin{figure}
\epsfxsize=7.25in
\epsfysize=4.5in
\epsfxsize=7.25in
\epsfysize=4.5in
\caption{Dependence of $Im\hat{G}^{T_3}$ with respect to $Im T_3$,
with the real part of $T_3$ fixed at $e^{\frac{i\pi}{6}}$ upper figure,
and at $1.3$ lower figure.}
\end{figure}
\newpage
\begin{figure}
\epsfxsize=7.25in
\epsfysize=4.5in
\epsfxsize=7.25in
\epsfysize=4.5in
\caption{Zero of $\hat{G}^{T_3}$ other than the fixed point. }
\end{figure}
\newpage
\begin{figure}
\epsfxsize=7.2in
\epsfysize=5.0in
\caption{Dependence of $V_{eff}$ on the $U_3$ modulus for fixed values of the
other moduli at their minimum see (\ref{photia}).}
\end{figure}
\newpage
\begin{figure}
\epsfxsize=6in
\epsfysize=8in
\caption{The solution on the unit circle for the $T_1$ modulus}
\end{figure}
\newpage
\begin{figure}
\epsfxsize=6in
\epsfysize=8in
\caption{Variation of $Re T$ at the minimum of $V_{eff}$ with
$\rho$, in multiple gaugino condensate models.}
\end{figure}
\newpage
\begin{figure}
\epsfxsize=6in
\epsfysize=4in
\caption{Variation of the logarithm of $Im \hat{G}$ with respect to $Re T$.}
\end{figure}
\newpage
\begin{figure}
\epsfxsize=6in
\epsfysize=6in
\caption{The effective potential for the non-perturbative
dilaton K$\rm{\ddot{a}}$hler potential in the overall modulus case.}
\end{figure}
\newpage
\begin{figure}
\epsfxsize=6in
\epsfysize=8in
\caption{Solutions of $V_{eff}$  
inside the fundamental domain with $m=n=1$, in the overall single modulus
case.}
\end{figure}
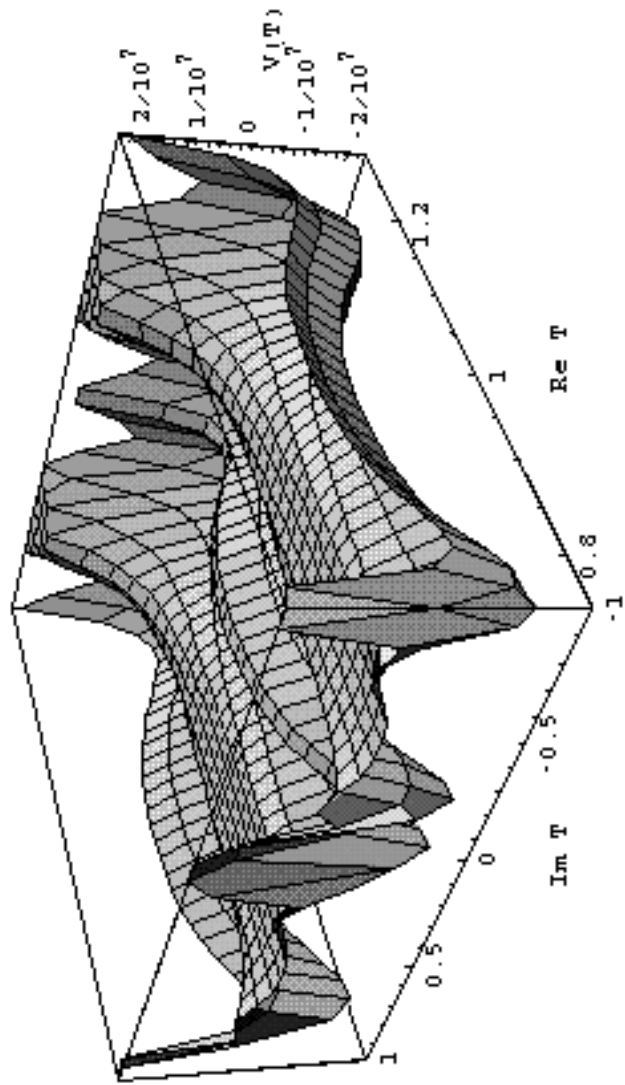
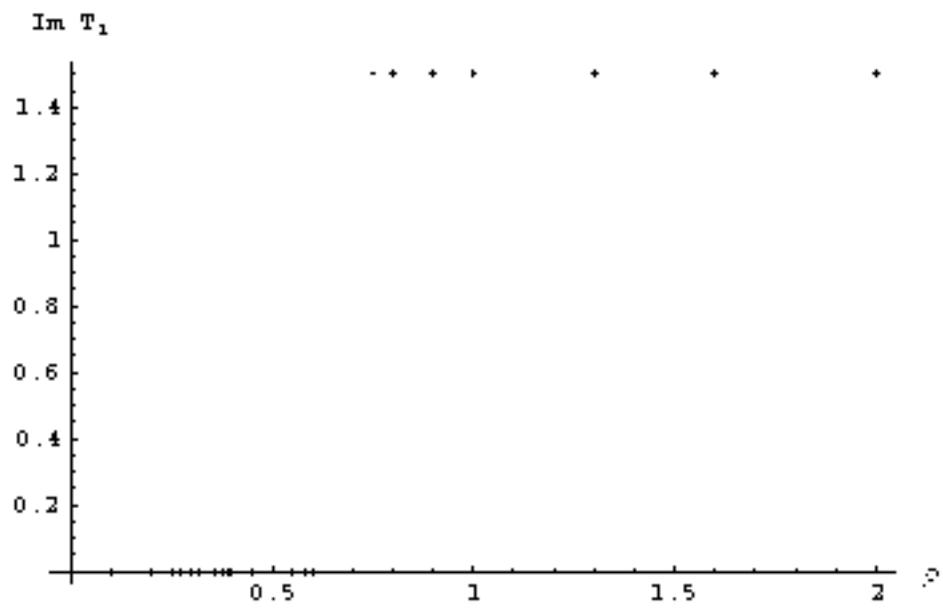
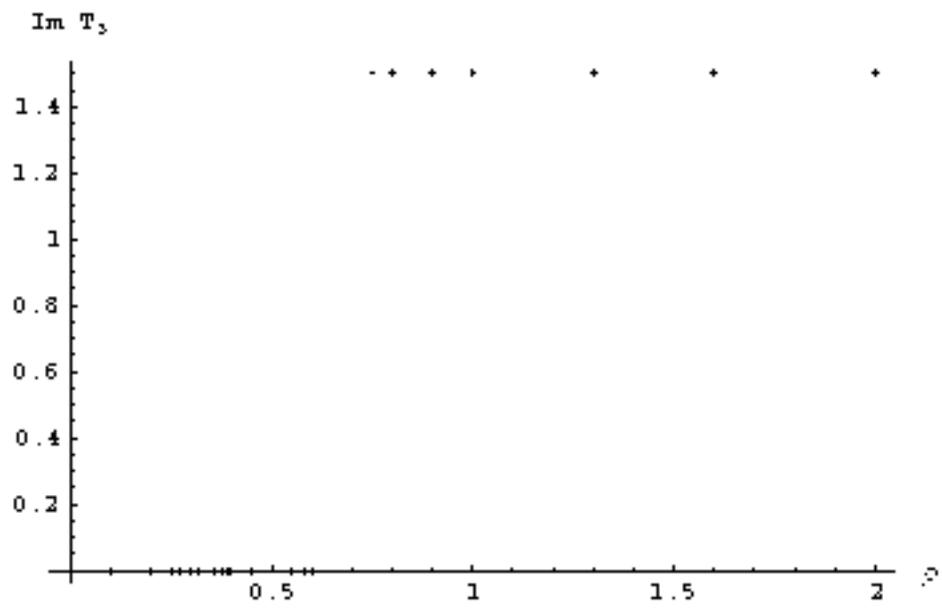
\end{document}